\documentclass[10pt,journal,compsoc]{IEEEtran}
\usepackage[table]{xcolor}

\usepackage{pifont}
\newcommand{\checkmark}{\ding{51}}%
\newcommand{\xmark}{\ding{55}}%
\usepackage{listings}

\definecolor{mGreen}{rgb}{0,0.6,0}
\definecolor{mGray}{rgb}{0.5,0.5,0.5}
\definecolor{mPurple}{rgb}{0.58,0,0.82}
\definecolor{backgroundColour}{rgb}{0.95,0.95,0.92}

\lstdefinestyle{CStyle}{
    backgroundcolor=\color{backgroundColour},   
    commentstyle=\color{mGreen},
    keywordstyle=\color{magenta},
    numberstyle=\tiny\color{mGray},
    stringstyle=\color{mPurple},
    basicstyle=\footnotesize,
    breakatwhitespace=false,         
    breaklines=true,                 
    captionpos=b,                    
    keepspaces=true,                 
    numbers=left,                    
    numbersep=5pt,                  
    showspaces=false,                
    showstringspaces=false,
    showtabs=false,                  
    tabsize=2,
    language=C
}
\usepackage{amsmath}
\usepackage{textcomp}
\usepackage{listings}
\usepackage{color}
\usepackage{threeparttable}
\definecolor{dkgreen}{rgb}{0,0.6,0}
\definecolor{gray}{rgb}{0.5,0.5,0.5}
\definecolor{mauve}{rgb}{0.58,0,0.82}

\ifCLASSOPTIONcompsoc
  \usepackage[nocompress]{cite}
\else
  \usepackage{cite}
\fi

\ifCLASSINFOpdf
   \usepackage[pdftex]{graphicx}
   \DeclareGraphicsExtensions{.pdf,.jpeg,.png}
\else
\fi

\usepackage{array}
\usepackage{booktabs}
\usepackage{multirow}
\usepackage{subcaption}

\newcommand{\ignore}[1]{}

\usepackage{float}

\usepackage{url}
\usepackage{enumitem}
\usepackage{xcolor}
\usepackage{soul}
\usepackage{amsmath}
\usepackage{algorithmic}
\usepackage[ruled,vlined]{algorithm2e}
\makeatletter
\def\BState{\State\hskip-\ALG@thistlm}
\makeatother
\sethlcolor{black}
\makeatletter
\newif\if@blind
\@blindfalse %\@blindtrue %use \@blindfalse on final version
\if@blind \sethlcolor{black}\else

\fi

\begin{document}
\title{MGPU-TSM: A Multi-GPU System with\\ Truly Shared Memory}

\vspace{-0.1in}
%\author{\IEEEauthorblockN{yyy}\\ %%
%  \IEEEauthorblockA{
%    \{zzz\}@bu.edu} \vspace{-0.15in}}

\author{
	Saiful~A.~Mojumder$^1$,
	Yifan~Sun$^2$,
	Leila~Delshadtehrani$^1$,
	Yenai Ma$^1$,
	Trinayan Baruah$^2$,\\
	\normalfont {Jos\'e~L.~Abell\'an$^3$,
	John~Kim$^4$,
	David~Kaeli$^2$,
	Ajay~Joshi$^1$}\\
	\normalfont{\small $^1$ECE Department, Boston University;}
	\normalfont{\small $^2$ECE Department, Northeastern University;}\\
	\normalfont{\small $^3$CS Department, UCAM;}
	\normalfont{\small $^4$School of EE, KAIST;}\\
	\normalfont{\small \{msam, delshad, yenai, joshi\}@bu.edu, \{yifansun, tbaruah, kaeli\}@ece.neu.edu,}\\
  \normalfont{\small jlabellan@ucam.edu, jjk12@kaist.edu}
  \vspace{-3mm}
  }
%\markboth{Journal of Computer Architecture Letters,~Vol.~16, No.~2, July-December~2017}%
%{Shell \MakeLowercase{\textit{et al.}}: Nile: A Programmable Hardware Monitor}

%\IEEEpubid{\begin{minipage}{\textwidth}\ \\[12pt]
%    Manuscript received 03 Nov. 2017; accepted 06 Dec. 2017. \\
%    Date of publication 0 . 0000; date of current version 0 . 0000.
%    \end{minipage}}
\IEEEtitleabstractindextext{%
\begin{abstract}

The sizes of GPU applications are rapidly growing. They are exhausting the compute and memory resources of a single GPU, and are demanding the move to multiple GPUs. However, the performance of these applications scales sub-linearly with GPU count because of the overhead of data movement across multiple GPUs. Moreover, a lack of hardware support for coherency exacerbates the problem because a programmer must either replicate the data across GPUs or fetch the remote data using high-overhead off-chip links. To address these problems, we propose a multi-GPU system with truly shared memory (MGPU-TSM), where the main memory is physically shared across all the GPUs. We eliminate remote accesses and avoid data replication using an MGPU-TSM system, which simplifies the memory hierarchy. Our preliminary analysis shows that MGPU-TSM with 4 GPUs  performs, on average, 3.9$\times$ better than the current best performing multi-GPU configuration for standard application benchmarks. 
\end{abstract}
\vspace{-0.05in}
\begin{IEEEkeywords}
Multi-GPU, Shared Memory, RDMA \vspace{-0.05in}
\end{IEEEkeywords}}

\maketitle

\IEEEdisplaynontitleabstractindextext

\IEEEpeerreviewmaketitle

\vspace{-0.2in}
\section{Introduction}
\label{sec:Intro}
\vspace{-0.05in}
\begin{table*}[!htbp]
\caption{Comparison of different communication mechanisms available in existing MGPUs vs. the communication scheme in MGPU-TSM. We compare the programmability and memory usage of each mechanism w.r.t. P2P Memcpy. Latency and BW is compared w.r.t. local MM access latency and BW. `\xmark', `\checkmark', and `\checkmark \checkmark' indicate  `no',  `fair' , and `good', respectively . 
\vspace{-0.1in}}
\label{tab:communication}
\resizebox{\textwidth}{!}{%
\begin{tabular}{@{}lllllcc@{}}
\toprule
Method & Definition  & \begin{tabular}[c]{@{}l@{}}MM Access \\ Latency\end{tabular}  & \begin{tabular}[c]{@{}l@{}}MM Access \\ Bandwidth\end{tabular} & \begin{tabular}[c]{@{}l@{}}Data \\ Duplication\end{tabular} & \begin{tabular}[c]{@{}l@{}}Improves\\ Programmability\end{tabular} & \begin{tabular}[c]{@{}l@{}}Improves GPU\\ Mem. Usage\end{tabular} \\ \midrule
P2P Memcpy     & \begin{tabular}[c]{@{}l@{}}Data copy from one GPU MM to another GPU MM\end{tabular}                                            & \begin{tabular}[c]{@{}l@{}}High \end{tabular} & Low & Yes                                                    & --                                             & --                                              \\ \hline
P2P Direct     & \begin{tabular}[c]{@{}l@{}}Data is accessed directly from the remote GPU memory \\ and cached in the requesting GPU's L1\$\end{tabular}                         & \begin{tabular}[c]{@{}l@{}}High\end{tabular} & Low &  Partial                                                & \checkmark                                          & \checkmark                                           \\ \hline
Zerocopy       & \begin{tabular}[c]{@{}l@{}}Data is directly accessed from CPU memory by all GPUs\\ without copying the data into GPU memory or GPU cache\end{tabular}               & \begin{tabular}[c]{@{}l@{}}Extremely\\ high\\\end{tabular} & Low &  No                                                & \checkmark \checkmark                                         & \xmark                                                \\ \hline
Unified Memory & \begin{tabular}[c]{@{}l@{}}Data is either transferred or accessed directly from the current \\ owner based on how the runtime decides to serve a page fault\end{tabular} & \begin{tabular}[c]{@{}l@{}} Extremely\\ High\end{tabular}  & Low  &      No                                         & \checkmark \checkmark                                         & \checkmark       
 \\ \hline
MGPU-TSM & \begin{tabular}[c]{@{}l@{}}All CPUs and GPUs can access the physically shared main \\ memory seamlessly using a low latency network \end{tabular} & \begin{tabular}[c]{@{}l@{}}Low\\ \end{tabular}    & High       & No                                                & \checkmark \checkmark                                        & \checkmark \checkmark     
\\ \bottomrule
\end{tabular}%
}
\vspace{-0.2in}
\end{table*}
Graphics processing units (GPUs) have become the system of choice for accelerating a variety of workloads including deep learning, graph applications, data mining, and big data processing. The size of these applications is growing continuously, and these applications are exhausting the compute and memory resources in single-GPU systems. Hence, the community is actively migrating towards using multi-GPU (MGPU) systems to accelerate the above-mentioned workloads. To enable inter-GPU communication, GPU vendors have proposed a number of mechanisms (see Table~\ref{tab:communication}).  However, achieving near-ideal speedup (w.r.t. a single GPU) when using multiple GPUs is challenging because of the inefficiencies in MGPU system design and the associated programming model.   

\textbf{Inefficiency 1:} In the existing discrete MGPU systems, each GPU has its own local main memory (MM) as shown in Figure~\ref{fig:tsm}(left). Each GPU in the MGPU system can access the other GPUs' MM through low-bandwidth high-latency links. These off-chip links have 5$\times$ to 10$\times$ lower bandwidth (BW) (for transferring data between GPUs, and between CPU and GPU) than the BW for accessing local MM of a GPU. Thus, accessing a remote GPU's MM increases the application execution time. Moreover, we observe non-uniform memory access (NUMA) effects when accessing a remote memory resulting in  under-utilization of GPU compute resources, and therefore sub-optimal performance.

\textbf{Inefficiency 2:}  Today's MGPU programming model requires a programmer to manually maintain coherency by replicating data and/or accessing non-cached data from a remote memory using the expensive off-chip links. As a result, there is additional traffic traversing through the off-chip links. In addition, the existing weak data-race-free (DRF) consistency model for GPUs requires additional efforts from the programmer to avoid data races by providing explicit barriers. 

As a result of these inefficiencies, we cannot leverage the full potential of MGPU systems. We provide more details about these two inefficiencies with experimental evaluation in Section~\ref{sec:motiv}. Researchers have proposed various solutions to address the aforementioned inefficiencies in the MGPU systems. In particular, the solution with identical objectives to ours was by Arunkumar et al. \cite{arunkumar2017mcm}, who proposed a package-level integration of multi-chip-module GPU (MCM-GPU) (see Figure~\ref{fig:tsm}(left)), where each GPU module has its own local DRAM. Here local accesses have low-latency, but remote accesses have very high latency. In parallel, other hardware and software optimizations such as L1.5\$~\cite{arunkumar2017mcm}, CARVE~\cite{young2018combining}, and HMG~\cite{hmg} have been proposed to address the two inefficiencies mentioned earlier. 

\ignore{

Today's multi-GPU systems lack hardware support for coherency and require programmer to maintain coherency manually. GPUs support data-race-free (DRF) weak consistency model. As a result, programming a GPU is a challenging task as a programmer has to use barriers, replicate data or leverage atomicity to manually maintain coherency and consistency. When multiple GPUs are used to accelerate an application, maintaining coherency and consistency both within a GPU and among the GPUs is an even more challenging task.

To provide communication across multiple GPUs' memory, a number of methods have been introduced by GPU vendors (see  Table~\ref{tab:communication}). These mechanisms provide the programmer the option to improve programmability or performance or memory utilization efficiency. 

Previously, GPUs needed to communicate with each other with the help of a host CPU. With the introduction of P2P memcpy, GPUs can directly copy to/from another GPU's MM; therefore, P2P memcopy provides a faster means of communication. However, using this method the programmer must manually control the data transfer while writing the program code and thus the programmer faces additional complexity in programming GPUs. Moreover, P2P memcpy creates multiple copies of data and leads to wastage of scarce GPU memory. To alleviate these issues, P2P direct access (known as remote data memory access, RDMA) has been introduced. RDMA allows one GPU to fetch data from another GPU (main memory or cache) to its own cache and use it directly without copying into its main memory. A single pointer can be used to access data across all the GPUs, which reduces the burden of the programmer. To handle the GPU memory capacity issues, Zerocopy method allows all the GPUs to use pinned CPU memory directly (no copying is allowed in GPU cache or main memory). However, Zerocopy is known to introduce high access latency data access as well as serialization in data access if multiple GPUs try to access the same memory region. Unified memory (UM) is an alternate approach to address the memory capacity issues as well as reduce programmer effort. Using this method, data is user transparently transferred across the GPUs with the help of page fault mechanism. However, UM has been shown to degrade performance because of high latency in serving the page faults using the slow off-chip links. }

To simplify programming, reduce the data transfer latency and increase the memory utilization efficiency, we propose a multi-GPU system with truly shared memory (MGPU-TSM). Unlike the MCM-GPU (see Figure~\ref{fig:tsm}), an MGPU-TSM system allows all GPUs to directly access the entire physical main memory of the system, thus eliminating  non-uniform memory access (NUMA) effects observed in traditional MGPU systems. In addition, an MGPU-TSM does not require L1.5\$ to reduce remote access overhead. Moreover, MGPU-TSM paves the way to accommodate low-overhead coherence protocol as well as a simpler consistency model for MGPU systems. In this work, we compare the performance of an MGPU-TSM design with state-of-the-art RDMA-- and unified memory (UM)--based MGPU designs using MGPUSim~\cite{sun2019mgpusim} to demonstrate the benefits of MGPU-TSM systems. 

\ignore{
\begin{figure}[!htp]
\centering
   \begin{subfigure}[b]{0.45\textwidth}
   \includegraphics[width=\textwidth]{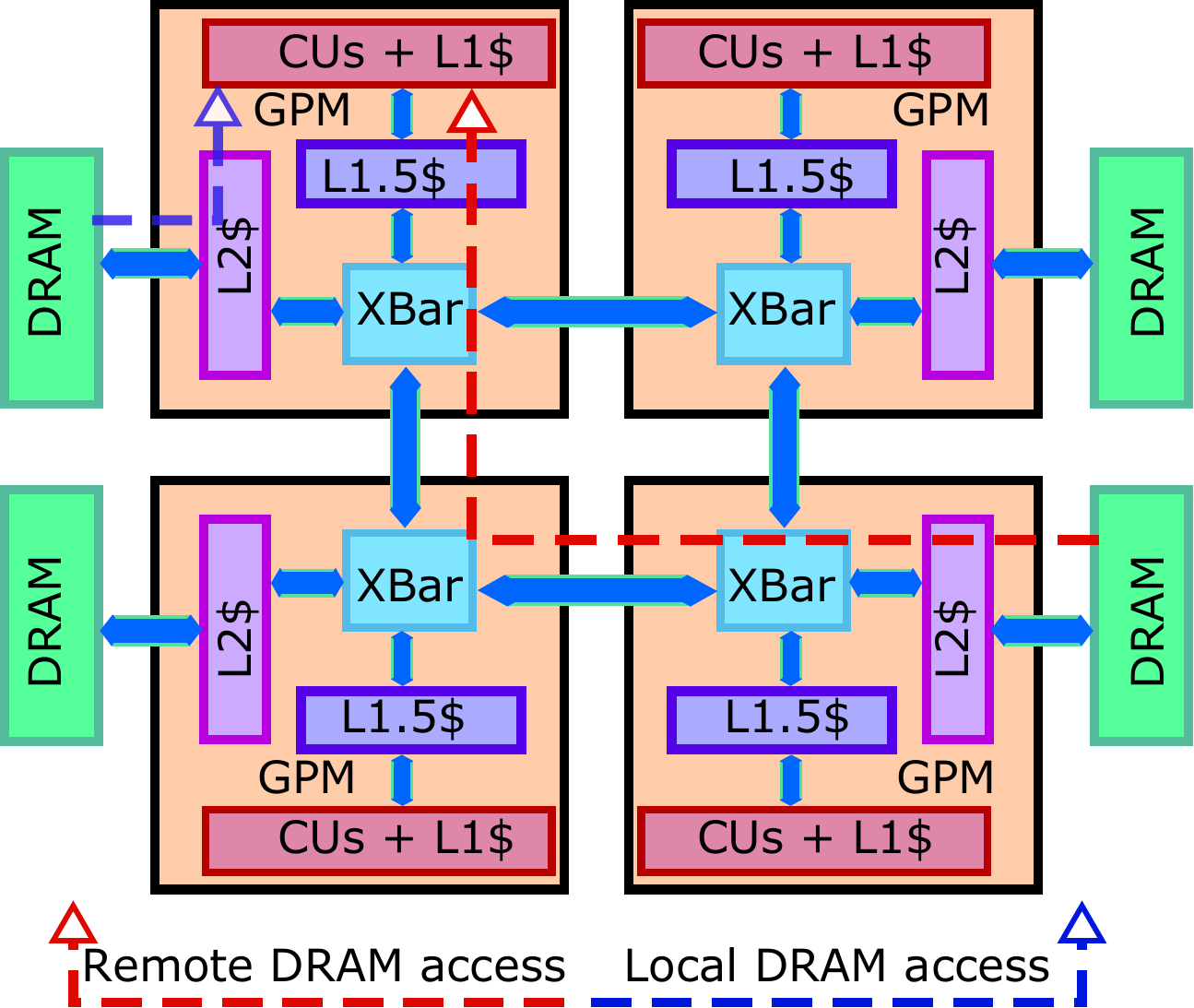}
   \caption{}
   \label{fig:Ng1} 
\end{subfigure}

\begin{subfigure}[b]{0.4\textwidth}
   \includegraphics[width=\textwidth]{tsm.pdf}
   \caption{}
   \label{fig:Ng2}
\end{subfigure}

\caption{(a) MCM-GPU (b) Proposed MGPU-TSM.\vspace{-0.25in}}

\end{figure}
\vspace{-0.15in}

\begin{figure}[]
  \centering
    \subfloat{
   \includegraphics[width=0.43\columnwidth]{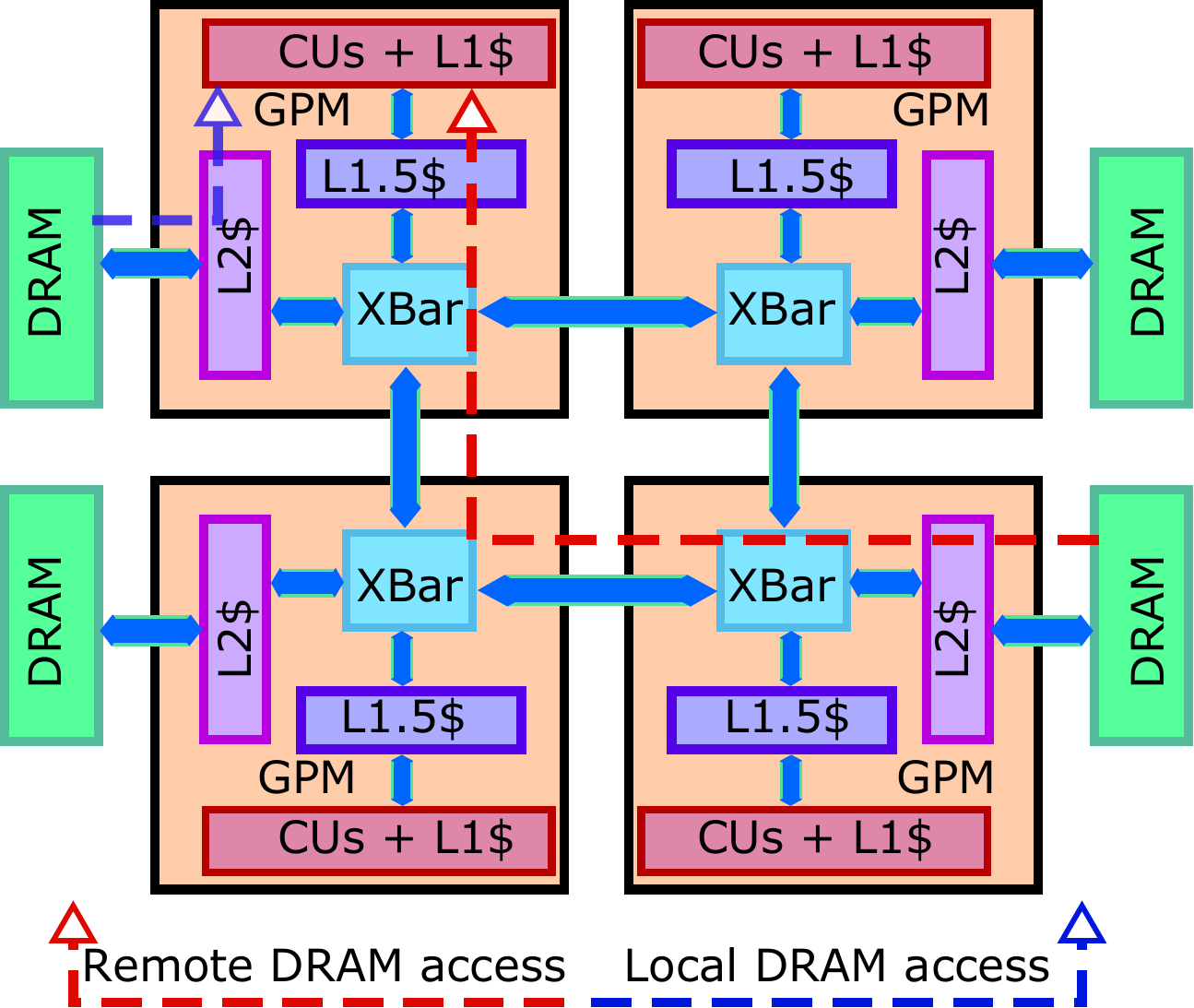}
   \label{fig:Ng1} 
  }
  \subfloat{
   \includegraphics[width=0.57\columnwidth]{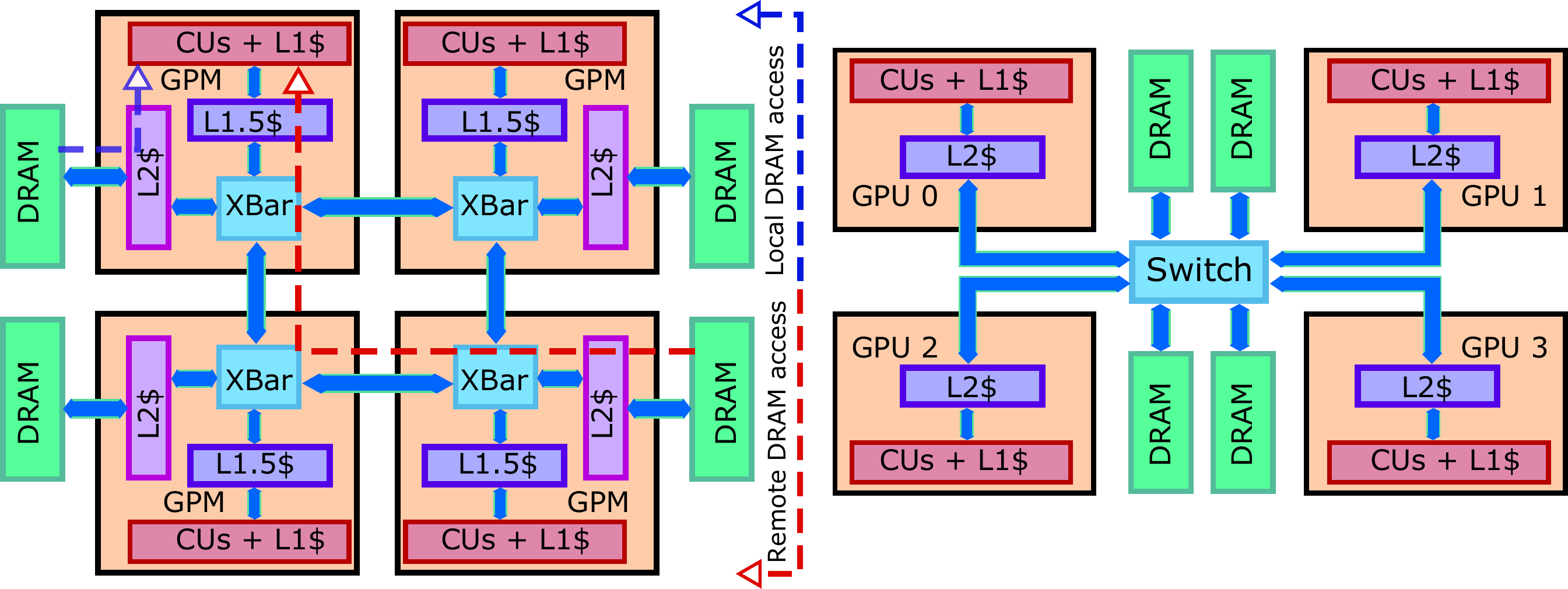}
   \label{fig:Ng2}
  }
\caption{MCM-GPU system (left) and Proposed MGPU-TSM system (right).\vspace{-0.35in}}
\end{figure}}

\begin{figure}[t]
  \centering
  \includegraphics[width=\columnwidth]{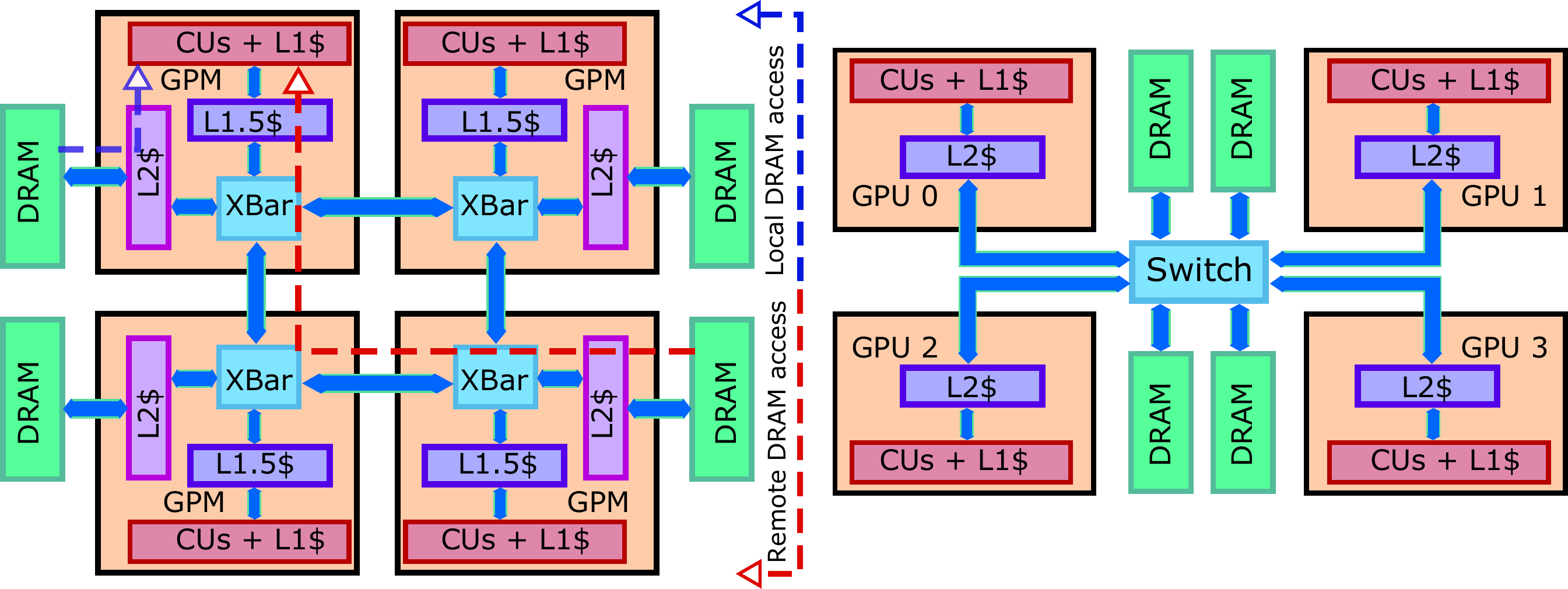}
\caption{\textit{MCM-GPU system (left) and Proposed MGPU-TSM system (right).}\vspace{-0.1in}}
  \label{fig:tsm}
\end{figure}
%\vspace{-0.2in}
\section{Challenges in Existing MGPU Systems}\label{sec:motiv}

\subsection{RDMA Access Cost}
In this section, using the data access latency metric, we present the motivation for providing shared main memory in an MGPU system. Here, we run the commonly-used matrix multiplication kernel \texttt{SGEMM}, from NVIDIA's cuBLAS library~\cite{nvidia2008cublas}, on an MGPU system with V100 GPUs (compute capability of 7.0). We use two GPUs connected through NVLink 2.0 (50 GB/s bidirectional bandwidth). The conclusions of our analysis should be broadly applicable to systems with more than 2 GPUs that use GPU-GPU RDMA.

The computations in the \texttt{SGEMM} kernel consist of three matrices A, B, and C. In our experiment, we distribute the matrices in the memory of two GPUs (GPU0 and GPU1) and examine the performance degradation caused by different degrees of remote access (using P2P direct access as an example) when the \texttt{SGEMM} is executed on GPU0. We use the \texttt{aL-bR} format to represent \texttt{a}\% local access and \texttt{b}\% remote access for GPU0, where \texttt{a} and \texttt{b} are integers. We evaluate the following four matrix distributions across memory:

\begin{enumerate}
    \item Matrices A, B and C are in GPU0's memory. This leads to 100\% local access for GPU0 (\texttt{100L-0R}).
    \item Matrices A and B are in GPU0's memory, and C is in GPU1's memory (\texttt{67L-33R}).
    \item Matrix A is in GPU0's memory, and matrices B and C are in GPU1's memory (\texttt{33L-67R}).
    \item Matrices A, B and C are in GPU1's memory. This leads to 100\% remote access for GPU0 (\texttt{0L-100R)}.
\end{enumerate}

\begin{figure}[t]
  \centering
  \includegraphics[width=\columnwidth]{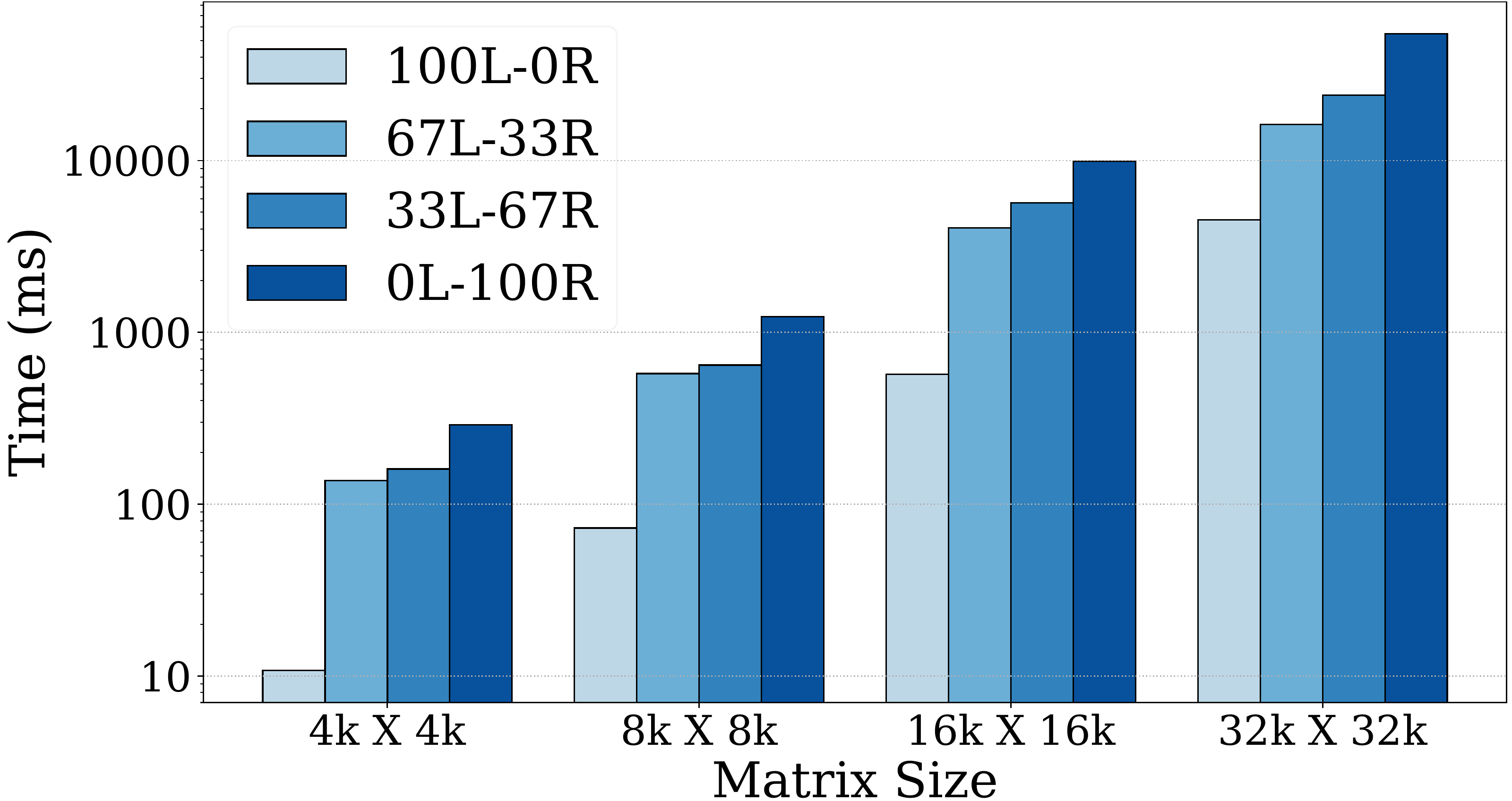}
  \caption{\textit{Runtime of \texttt{SGEMM} kernel from cuBLAS library for different matrix sizes. Each bar corresponds to a different distribution of local and remote memory accesses.}}
  \label{fig:mot}
  \vspace{-0.25in}
\end{figure}

Figure~\ref{fig:mot} shows the runtime for the \texttt{SGEMM} kernel execution with different matrix sizes for the above four matrix distributions. For smaller matrix sizes, accessing remote memory is very expensive because of the fixed remote access overhead. The runtime of \texttt{SGEMM} for the \texttt{0L-100R} distribution for a 4k$\times$4k matrix is 27$\times$ longer than that of the \texttt{100L-0R} distribution. On the other hand, the runtime of \texttt{SGEMM} for the \texttt{0L-100R} distribution for the 32k$\times$32k matrix is 12.2$\times$ longer than that of the \texttt{100L-0R} distribution. Here, the fixed remote access overhead gets amortized. From these experiments, we can see the significant impact of remote accesses on performance, and in turn, argue that to improve the performance of applications, we need to avoid remote accesses as much as possible.
%\vspace{-0.5in}

%\vspace{-0.15in}
\subsection{Data Sharing and Programmability}\label{sec:prog}
Data sharing across multiple GPUs during kernel execution leads to programming challenges, as the programmer must choose between programmability and performance. In this section, we examine the DNN training process on MGPU systems, when leveraging different data-parallelism schemes. We highlight how different mechanisms trade-off programmability for performance. The three stages of DNN training include forward propagation (FP), backward propagation (BP), and weight update (WU). During the FP and BP stages, different GPUs calculate their local stochastic gradient descents (SGDs) that are later used to update the values of weights used for the next iteration. \footnote{More details about the training stages can be found in~\cite{mojumder2018profiling}.}  

In Algorithms~\ref{algo:memcpy}, ~\ref{algo:p2p} and~\ref{algo:um}, we consider three different ways a programmer can perform the WU stage. We will assume a 2-GPU MGPU system here. Algorithm~\ref{algo:memcpy} shows that when using memcpy, the programmer must maintain coherence explicitly by periodically copying data to GPU1's memory. Thus, there is an additional copy of data i.e. SGD (\texttt{gGPU1}) in GPU0's memory, leading to additional memory usage. Nonetheless, this mechanism can be efficient in terms of kernel runtime because P2P memcpy can run asynchronously. Algorithm~\ref{algo:p2p} shows how P2P direct access with RDMA can eliminate the data copy step, but at the expense of accessing data using off-chip links. Still, the programmer must transfer the data from the CPU to the GPUs. Algorithm~\ref{algo:um} illustrates that a shared main memory could ease programmability and eliminate explicit GPU-to-GPU or CPU-to-GPU data transfers. Note that UM and Zerocopy solutions use Algorithm~\ref{algo:um}. UM, as proposed by NVIDIA, eases programming with a software abstraction, but suffers from performance degradation due to inefficient page-fault support and expensive remote accesses~\cite{trinayan-hpca2020}. A Zerocopy solution does not use GPU memory at all. The GPUs access {\em pinned} CPU memory using the off-chip (PCIe) links~\cite{negrut2014unified}. We argue that we need a solution which would not trade-off programmability to gain performance. A programmer can use Algorithm 3 on our envisioned MGPU-TSM and enjoy both ease of programmability and high performance. 

\vspace{-0.1in}
\begin{algorithm}[h] \SetAlgoLined \small
 Initialization: \texttt{weights} in CPU \; 
 Copy \texttt{weights} from CPU to GPU0 $\rightarrow$ \texttt{wGPU0} \ ;\\ 
 Copy \texttt{weights} from CPU to GPU1 $\rightarrow$ \texttt{wGPU1} \ ;\\
 FP+BP on GPU0 using \texttt{wGPU0} $\rightarrow$ \texttt{gGPU0} \ ;\\ 
 FP+BP on GPU1 using \texttt{wGPU1} $\rightarrow$ \texttt{gGPU1} \ ;\\ 
 Copy \texttt{gGPU1} from GPU1 to GPU0  $\rightarrow$ \texttt{gGPU0Copy} \ ;\\
 WU on GPU0 using \texttt{(gGPU0,gGPU0Copy)} $\rightarrow$ \texttt{wGPU0} \ ;\\ 
Copy \texttt{wGPU0} from GPU0 to GPU1 $\rightarrow$ \texttt{wGPU1} \ ;\\
\caption{Using Memcpy$^*$}
\label{algo:memcpy}

\end{algorithm}
\vspace{-0.3in}
\begin{algorithm}[h] \SetAlgoLined \small
 Initialization: \texttt{weights} in CPU \;
 Copy \texttt{weights} from CPU to GPU0 $\rightarrow$ \texttt{wGPU} \ ;\\ 
 FP+BP on GPU0 using \texttt{wGPU} $\rightarrow$ \texttt{gGPU0} \ ;\\ 
 FP+BP on GPU1 using \texttt{wGPU} $\rightarrow$ \texttt{gGPU1} \ ;\\
 WU on GPU0 using \texttt{(gGPU0,gGPU1)} $\rightarrow$ \texttt{wGPU} \ ;\\
\caption{Using P2P direct access$^*$}
\label{algo:p2p}
\end{algorithm}
\vspace{-0.3in}
\begin{algorithm}[h] \SetAlgoLined \small
 Initialization: \texttt{weights} in CPU \; 
 FP+BP on GPU0 using \texttt{weights} $\rightarrow$ \texttt{g0} \ ;\\ 
 FP+BP on GPU1 using \texttt{weights} $\rightarrow$ \texttt{g1} \ ;\\
 WU on GPU0 using \texttt{(g0, g1)} $\rightarrow$ \texttt{weights} \ ;\\ 
\caption{Using shared main memory$^*$}
\label{algo:um}
\end{algorithm}
\vspace{-0.2in}
$^*$In the pseudocode, the right arrows point to destination variables of an operation.

%\vspace{-0.15in}
\section{MGPU-TSM System}
\ignore{
\begin{figure}[t]
  \centering
  \includegraphics[width=\columnwidth]{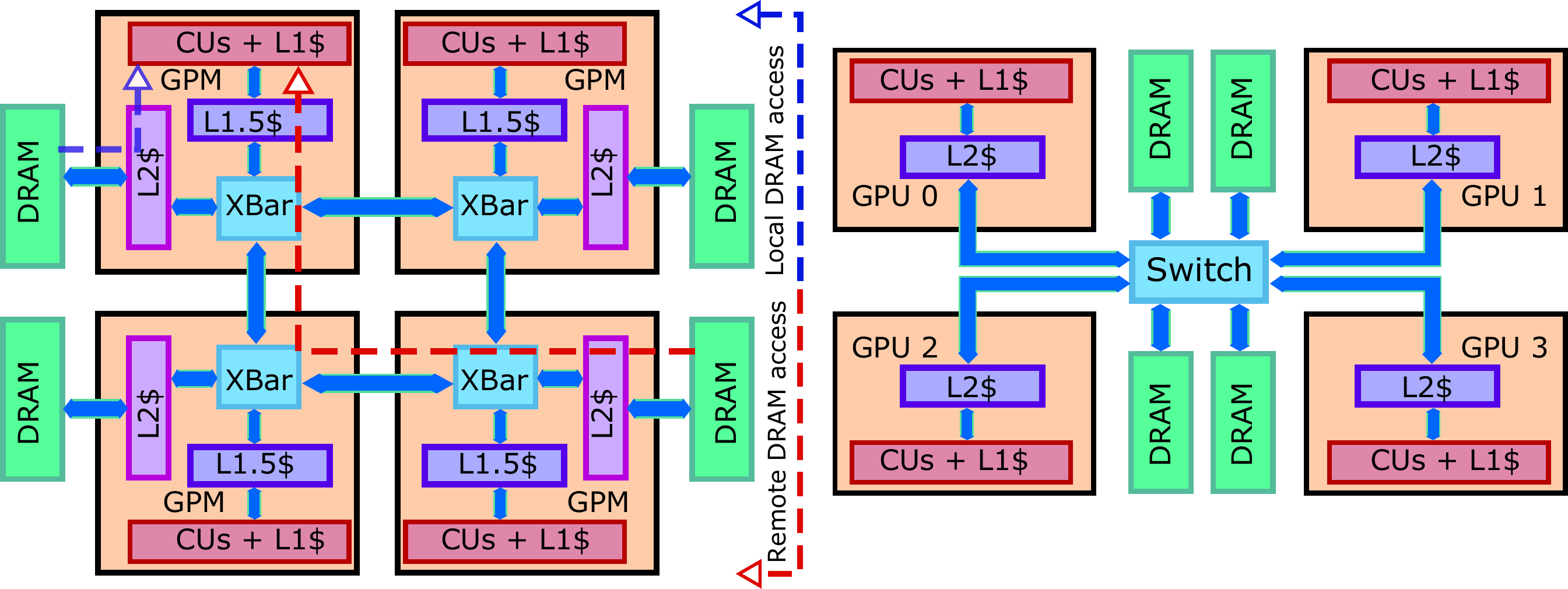}
  \caption{Simplified diagram for the TSM system}
  \label{fig:tsm}
  %\vspace{-0.25in}
\end{figure}
}
To explain and evaluate our envisioned MGPU-TSM architecture, we consider an MGPU-TSM system consisting of 4 GPUs, 1 CPU and 4 HBM stacks that provide a total of 32GB MM (we are using a 32GB capacity as an example to explain the MGPU-TSM architecture -- our MGPU-TSM system works with larger memory). The specifications of the GPU, CPU, and HBM stacks are provided in Table~\ref{tab:component}. 

%\vspace{-0.15in}
\subsection{MGPU-TSM Architecture}
Figure~\ref{fig:tsm}(right) shows the logical view of our proposed MGPU-TSM system. We leverage the current common design for compute units (CUs), where each CU has a dedicated write-through L1\$. All the L1\$s are connected to the L2\$s using a crossbar network. For our proposed MGPU-TSM system, we make changes to the memory hierarchy, starting from L2\$ down to the MM. 

GPUs typically have distributed L2\$ banks, where each L2\$ bank serves one memory controller (MC). In our envisioned MGPU-TSM system, we have 8 L2\$ banks per GPU and 4 HBM stacks that provide a total of 32 GB of MM. Thus, for each GPU, an L2 MC controls 4GB of memory. Each of the 8GB DRAMs is further distributed into 16 banks, where each bank has a 512MB capacity. 

Each L2 bank, as well as each DRAM bank, is connected to a centralized switch through a dedicated 32GB/s bidirectional link. Thus, each GPU has a total of 256GB/s of bidirectional BW between the L2\$ and MM. With 4 GPUs, the total BW is 1TB/s. This also implies that each memory access requires a two-hop communication, from L2\$ to the Switch, then from the Switch to MM, and vice versa. Recently, NVIDIA introduced NVSwitch~\cite{ishii2018nvswitch}, providing 18 ports and 928GB/s of bidirectional BW, supporting RDMA connectivity across multiple GPUs. Hence, our assumed 32-port switch with 1TB/s aggregate bidirectional BW is realistic.  

The key advantage of our TSM lies in physically-unified MM, providing uniform memory access (UMA) across the system. This physically-unified design completely removes the need for remote accesses. In addition, having a centralized location for data access by multiple GPUs provides the opportunity to coalesce data accesses at the MM level and makes it easier to provide support for coherency given the lower overhead in communication. Moreover, having more memory banks helps improve the throughput by an efficient allocation of data, i.e., allocating consecutive pages to neighboring DRAM banks in a round-robin manner.    

\begin{table}[t]
  \begin{center}
    \caption{Specification of MGPU-TSM components.\vspace{-0.07in}}
    \label{tab:component}
    \begin{tabular}{c|c|c|c|r}
     \toprule
    \textbf{Component} &  \textbf{Name} & \textbf{Tech. Node} & \textbf{Area} & \textbf{Power}\\
    &  & (nm) & (mm$^2$) & (W) \\
      \hline
    GPU &  RX 5700 & 7 & 151 & 180 \\
    CPU &  Ryzen 9 3950X   &  7 &  144$^*$ & 105\\
    Memory & HBM 2.0 & 14 & 92 & 21.4$^*$\\
        \bottomrule
    \end{tabular}
    \begin{tablenotes}
  \small
  \item $^*$ Determined using technology scaling rules.\vspace{-0.25in}
\end{tablenotes}
  \end{center}
\end{table}

%\vspace{-0.19in}
\subsection{Preliminary Evaluation}

\begin{table}[b]
  \centering 
  \vspace{-0.19in}
  \caption{GPU Architecture. \vspace{-0.15in}}
  \resizebox{0.48\textwidth}{!}{
  \begin{tabular}{@{}lll|llr@{}}
  \toprule

  \textbf{Component} & \textbf{Configuration} & \textbf{Count} &
  \textbf{Component} & \textbf{Configuration} & \textbf{Count}\\

  \textbf{per GPU} & &  & \textbf{per GPU}&\\ \midrule CU & 1.0 GHz & 32 & L1 Vector \$ & 16KB 4-way & 32\\

  L1 Scalar \$ & 16KB 4-way & 8 & L1I\$ & 32KB 4-way & 8\\

  L2\$ & 256KB 16-way & 8 & DRAM & 512MB HBM & 16 \\

  L1 TLB & 1 set, 32-way & 48 & L2 TLB & 32 sets, 16-way & 1\\

  \bottomrule
  \end{tabular}} % \vspace{-15pt}
  \label{tab:configuration}
  %\vspace{-0.25in}
\end{table}
In this section, we discuss the potential performance benefits of an MGPU-TSM system over the existing MGPU system configurations, i.e., MGPU systems that use RDMA P2P direct access (referred to as \texttt{RDMA}), and the MGPU system that uses unified memory (referred to as \texttt{UM}).  Table~\ref{tab:configuration} shows the configuration for each GPU in our evaluation, where we allocate memory by interleaving the pages across all the memory modules in the MGPU system. For a fair comparison, we use the same GPU specifications i.e. CU count, L1\$ and L2\$ sizes and number of total DRAM banks (16 for each GPU) for \texttt{RDMA}, \texttt{UM} and \texttt{TSM} configurations. We use a page size of 4KB. For the \texttt{RDMA} configuration, we use PCIe 4.0 links to provide 32GB/s bidirectional BW for remote accesses. \texttt{UM} provides a unified view of the total memory to the programmer by virtually combining the CPU and GPU memories. \texttt{UM} uses a first touch policy for page placement.   \ignore{When a device (CPU or GPU) requests data, a page containing the data is transferred to the memory of the requesting device based on first-touch policy. If another device requests data from the same page, a page-faults occurs. This page fault is typically served either by migrating the entire page to the second device or by \textcolor{red}{allowing non-cacheable direct access to the data.}} To evaluate our design we use the MGPUSim simulator~\cite{sun2019mgpusim}, which is designed specifically to support MGPU simulation. We use 12 standard benchmarks from the Hetero-Mark~\cite{heteromark}, PolyBench~\cite{pouchet2012polybench}, SHOC~\cite{shoc}, and DNNMark~\cite{dong2017dnnmark} benchmark suites for our preliminary evaluation. 

\begin{figure}[t]
  \centering
  \includegraphics[width=\columnwidth]{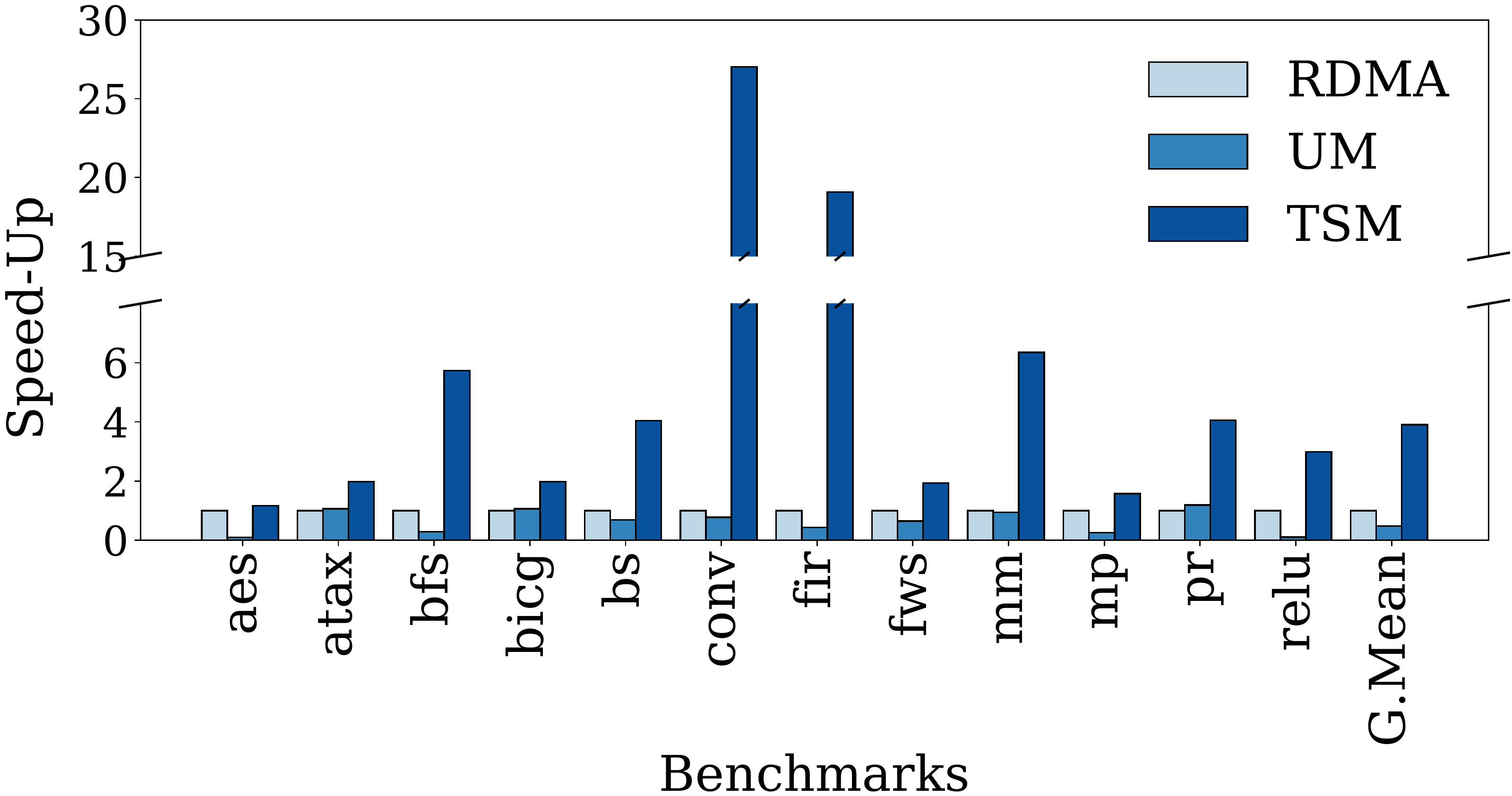}
  \caption{\textit{Speedup of proposed \texttt{TSM}, and  \texttt{UM}  w.r.t. \texttt{RDMA}.} }
  \label{fig:comp}
  \vspace{-0.15in}
\end{figure}

Figure~\ref{fig:comp} shows a comparison of \texttt{TSM}, \texttt{RDMA} and \texttt{UM}. \texttt{TSM} is, on average, 3.9$\times$ and 8.2$\times$ faster than \texttt{RDMA} and \texttt{UM}, respectively. \texttt{TSM} is faster than using \texttt{RDMA} because \texttt{RDMA} requires data copy operations between the CPU and GPUs. During kernel execution, all GPUs are required to use \texttt{RDMA} to access data residing on the other GPUs' memories. \texttt{UM} suffers from an expensive page fault service mechanism and page migration through the off-chip links.

%\vspace{-0.15in}
\section{MGPU-TSM System Design Challenges}
Our preliminary comparison of \texttt{TSM} with \texttt{RDMA} and \texttt{UM} shows that \texttt{TSM} is quite promising, but it also comes with several challenges. Here we discuss these challenges and our future research direction to address those challenges. 
%\vspace{-0.15in}
\subsection{Data Sharing Within and Across GPUs}
In the MGPU-TSM system, different CUs within and across GPUs can access the same memory location. Hence, we need a low-overhead scalable cache coherency and memory consistency model to maintain accuracy such as HALCONE~\cite{mojumder2020halcone}. Traditional snooping-based or directory-based coherency protocols, such as MESI and MOESI, can lead to large inter-GPU and intra-GPU communication latencies~\cite{singh2013cache}. Timestamp-based coherence~\cite{tabbakh2018g}, which allows auto-invalidation of cache blocks and reduces the traffic overhead, can be suitable for an MGPU-TSM system. A wide range of consistency models, including sequential consistency, weak consistency, and release consistency, have been proposed for single-GPU systems. We need to design consistency models for an MGPU-TSM system consisting of thousands of threads.
%\vspace{-0.15in}
\subsection{L2-to-MM Network}
The L2-to-MM network plays a critical role in the overall performance of an MGPU-TSM system. In our example system, we used direct links between L2 to the Switch and between the Switch to MM. As we scale the number of GPUs, the radix of the Switch grows proportionally. A high-radix switch leads to lower performance, and at the same time, the resulting area and power become problematic. In our future work, we will explore different high-BW low-latency networks that scale well with GPU count.
%\vspace{-0.15in}
\subsection{CPU-GPU Memory Accesses}
CPUs are typically latency-sensitive, while GPUs are BW-sensitive. Since the MGPU-TSM system provides the same physical memory to both CPUs and GPUs, it is imperative to design a network protocol that allows low-latency data access to the CPU and high-BW data access to the GPUs. 
%\vspace{-0.15in}
\subsection{Integration Technology}
To design a scaled-up MGPU-TSM system, we envision using 2.5D integration technology with multiple interposers. Each interposer will have multiple GPU chiplets, a CPU chiplet, and multiple HBM stacks. For intra-interposer communication, we can use electrical links, while for long-distance inter-interposer communication, we can use photonic links. To design such a multi-interposer system, we need to develop a cross-layer design automation technique that jointly optimizes the system architecture, circuit design, and physical design.

\ignore{
\begin{figure}[!htp]
\centering

\begin{subfigure}[b]{0.5\textwidth}
   \includegraphics[width=\textwidth]{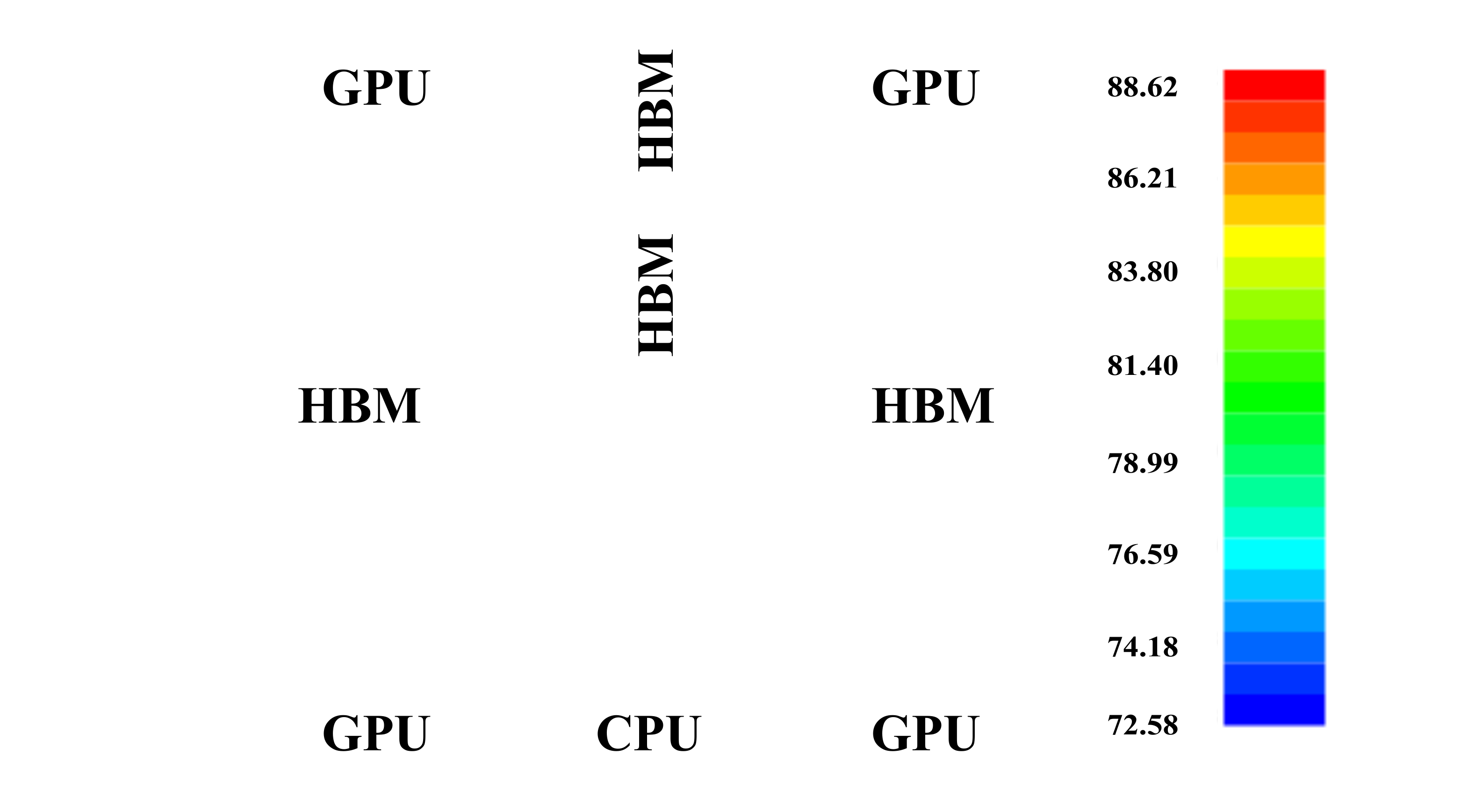}
   \caption{}
   \label{fig:Ng3}
\end{subfigure}

\begin{subfigure}[b]{0.5\textwidth}
   \includegraphics[width=\textwidth]{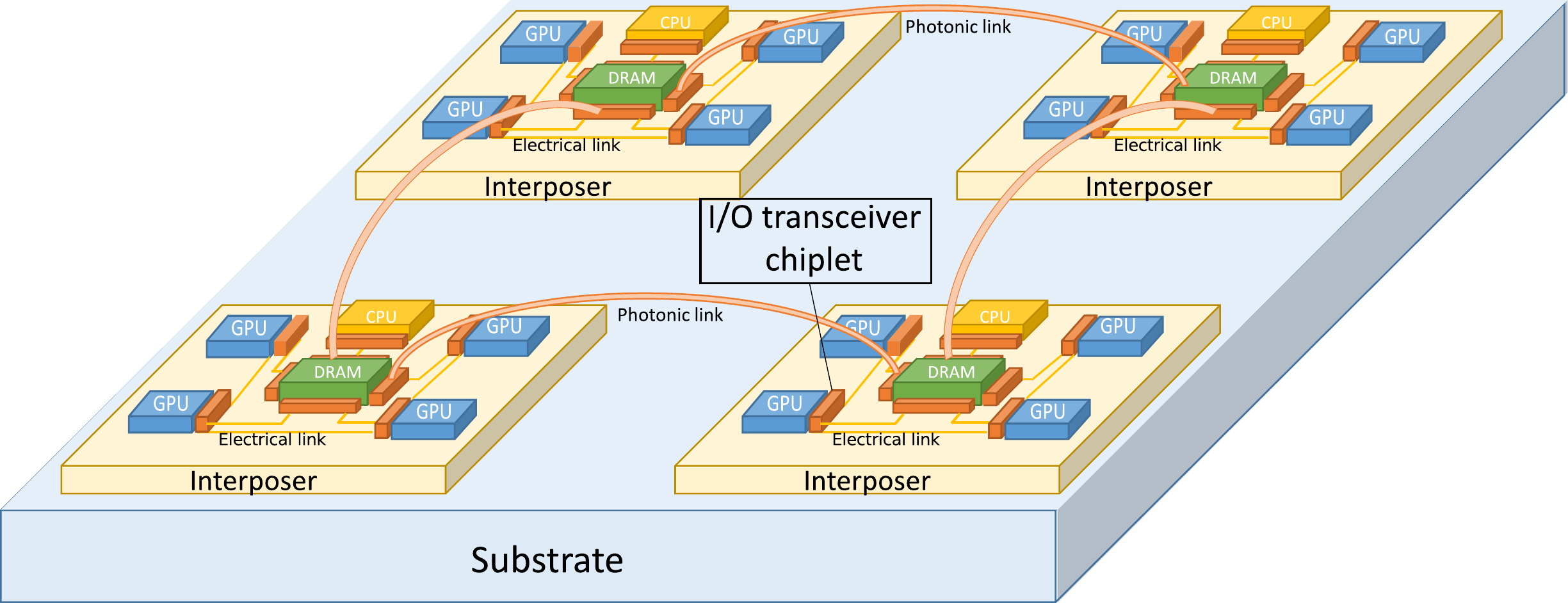}
   \caption{}
   \label{fig:Ng4}
\end{subfigure}
\caption{(a)  Thermal heatmap for MGPU-TSM (b) Scalable multi-interposer based multi-GPU system.}
\end{figure}
}
\ignore{
\begin{figure}[!htp]
  \centering
  \includegraphics[width=\columnwidth]{./figure/system_3Dview1.pdf}
  \caption{Multi-interposer based multi-GPU system}
  \label{fig:mcm}
  %\vspace{-0.25in}
\end{figure}
}

\ignore{
\begin{figure*}[ht] 
  \begin{subfigure}[b]{0.5\linewidth}
    \centering
    \includegraphics[width=\linewidth]{./figure/mcm.pdf} 
    \caption{Initial condition} 
    \label{fig7:a} 
    \vspace{4ex}
  \end{subfigure}%% 
  \begin{subfigure}[b]{0.5\linewidth}
    \centering
    \includegraphics[width=\linewidth]{./figure/tsm.pdf} 
    \caption{Rupture} 
    \label{fig7:b} 
    \vspace{4ex}
  \end{subfigure} 
  \begin{subfigure}[b]{0.5\linewidth}
    \centering
    \includegraphics[width=\linewidth]{./figure/final_heatmap.pdf} 
    \caption{DFT, Initial condition} 
    \label{fig7:c} 
  \end{subfigure}%%
  \begin{subfigure}[b]{0.5\linewidth}
    \centering
    \includegraphics[width=\linewidth]{./figure/system_3Dview1final_heatmap.pdf} 
    \caption{DFT, rupture} 
    \label{fig7:d} 
  \end{subfigure} 
  \caption{Illustration of various images}
  \label{fig7} 
\end{figure*}
}

%\input{thermal}
%\vspace{-0.15in}
\section{Conclusion}
\label{sec:conclusion}

In this work, we showed that the performance of MGPU systems is limited due to expensive remote data access through off-chip links. At the same time, programming MGPU systems is difficult due to a lack of hardware support for coherency. To address these issues, we propose an MGPU-TSM architecture that eliminates remote data access, improves memory utilization, and reduces programmer burden. We also highlight the major challenges we need to overcome to make MGPU-TSM viable.
%\vspace{-0.15in}

%\bibliographystyle{IEEEtran}
\bibliographystyle{abbrv-etal}
\bibliography{tsm-cal-2020.bib}

\end{document}